\documentclass[fleqn,usenatbib]{mnras}

\usepackage{newtxtext,newtxmath}

\usepackage[T1]{fontenc}
\usepackage{ae,aecompl}

\usepackage{graphicx}	

\newcommand{\hctsa}{\textit{hctsa}}

\newcommand{\gdor}{$\gamma$ Doradus}
\newcommand{\dsct}{$\delta$ Scuti}
\newcommand{\rrl}{RR~Lyr}
\newcommand{\kepler}{{\em Kepler}}
\newcommand{\tess}{{\em TESS}}
\newcommand{\kep}{\kepler}

\newcommand{\new}[1]{\textcolor{red}{\textbf{#1}}} 
\renewcommand{\new}[1]{#1} 

\title[Classifying \kepler\ light curves]{Classifying \kepler\ light curves for 12,000 A and F stars using supervised feature-based machine learning}

\author[Barbara et al.]{%
Nicholas H. Barbara,$^{1,2}$\thanks{E-mail: nicholas.barbara@sydney.edu.au}
Timothy R. Bedding,$^{1,2}$\thanks{E-mail: tim.bedding@sydney.edu.au}
Ben D. Fulcher,$^{1}$ \newauthor
Simon J. Murphy$^{1,2}$, and
Timothy Van Reeth$^{1,2,3}$
\\
$^1$Sydney Institute for Astronomy, School of Physics, University of Sydney 2006, Australia \\
$^2$Stellar Astrophysics Centre, Department of Physics and Astronomy, Aarhus University, Denmark\\
$^3$Institute of Astronomy, KU Leuven, Celestijnenlaan 200D, B-3001 Leuven, Belgium}

\date{Accepted XXX. Received YYY; in original form ZZZ}

\pubyear{2022}
\begin{document}
\label{firstpage}
\pagerange{\pageref{firstpage}--\pageref{lastpage}}
\maketitle

\begin{abstract}
With the availability of large-scale surveys like \kepler\ and \tess, there is a pressing need for automated methods to classify light curves according to known classes of variable stars.
We introduce a new algorithm for classifying light curves that compares 7000 time-series features to find those which most effectively classify a given set of light curves.
We apply our method to \kepler\ light curves for stars with effective temperatures in the range 6500--10,000\,K.  We show that the sample can be meaningfully represented in an interpretable five-dimensional feature space that separates seven major classes of light curves ($\delta$\,Scuti stars, $\gamma$\,Doradus stars, RR\,Lyrae stars, rotational variables, contact eclipsing binaries, detached eclipsing binaries, and non-variables).
We achieve a balanced classification accuracy of 82\% on an independent test set of \kepler\ stars using a Gaussian mixture model classifier.
We use our method to classify 12,000 \kepler\ light curves from Quarter 9 and provide a catalogue of the results. We further outline a confidence heuristic based on probability density with which to search our catalogue, and extract candidate lists of correctly-classified variable stars. 
\end{abstract}


%
%
\section{Introduction}

The use of machine learning is becoming increasingly common in astronomy \citep{Ball+Brunner2010,Graff++2014,Ivezic++2019-book,Baron2019}.  In particular, large-scale photometric surveys are producing light curves in numbers too large for humans to manually inspect and analyse.
Considerable efforts have gone into using machine learning to classify light curves from large ground-based surveys \citep[e.g.,][]{Carrasco-Davis++2019, Tsang++2019, Johnston++2019-classification, Cabral++2020, Hosenie++2020, Jamal+Bloom2020, szklenaretal2020, Bassi++2021, Zhang+Bloom2021}.  Such techniques have also been applied to light curves from NASA's \kepler\ and K2 missions \citep[e.g.,][]{Blomme++2010, Blomme++2011, Debosscher++2011, Bass+Borne2016, Armstrong++2016, Hon++2017,Hon++2018a,Hon++2018b, Johnston++2019-binaries, Kgoadi++2019-clustering, LeSaux++2019, Giles+Walkowicz2020, Kuszlewicz++2020, Audenaert++2021, Paul+Chattopadhyay2022}.

A range of algorithms have been proposed to classify light-curve databases according to known classes of stars, but these algorithms often rely on black-box machine-learning methods which limits their interpretability and hence ability to drive scientific understanding.
Those algorithms that are more interpretable rely on manually selected temporal or spectral features of light curves \citep[e.g.,][]{Pashchenko++2017}, with minimal comparison to the performance of alternatives from across a highly interdisciplinary time-series analysis literature.
Here we introduce a new algorithm for classifying light-curve databases that searches over 7000 time-series features to automatically find interpretable features relevant to classifying a given set of light curves.
\new{Our aim is to develop a simple, efficient classifier that uses these interpretable features to give us new insight into the classification of variable stars, while maintaining comparable performance with existing methods.}

Over the course of its four-year mission, the \kepler\ spacecraft collected light curves for nearly 200,000 stars, most of which show variability. Subsets of \kepler\ stars have been classified systematically, resulting in catalogues of about 2900 eclipsing binaries \citep{kirketal2016}, 16,000 oscillating red giants \citep[e.g.,][and references therein]{Yu++2018}, 2000 pulsating $\delta$~Scuti stars \citep{murphyetal2019}, and over 600 $\gamma$~Doradus pulsators \citep{lietal2020}.
Independently, \citet{Balona2018} used visual inspection of light curves and power spectra to classify over 20,000 A and F Kepler stars.  Finally, \citet{Audenaert++2021} have recently classified 167,000 light curves from one quarter of \kepler\ data (see Sec.~\ref{sec:audenaert}).

In this paper, we present our methods and provide classifications based on a single three-month quarter (Q9) for approximately 12,000 \kepler\ stars with effective temperatures in the range 6500--10,000\,K.
\new{Our primary interest is in pulsating stars and our chosen temperature range covers the classical instability strip, which has the richest variety of variability \citet[e.g.,][]{Kurtz2022}. There are relatively few stars in the Kepler sample that are hotter than this range, while pulsations on the cooler side are dominated by a single class (solar-like oscillations) that have already been extensively classified studied (see \citealt{Jackiewicz2021} for a recent review).}
We note that we have previously used our method to identify samples of \dsct\ stars \citep{murphyetal2020a} and \gdor\ stars \citep{lietal2020}.
\new{Finally, we compare our classification of \kepler\ stars to labels assigned by a less interpretable, performance-focused classifier in \cite{Audenaert++2021}, where we find similar results.}

%
%
\section{Training data}

\subsection{A Selection of Variable Stars} 
\label{ssec:classes}

\begin{figure*}
\includegraphics[width=\linewidth]{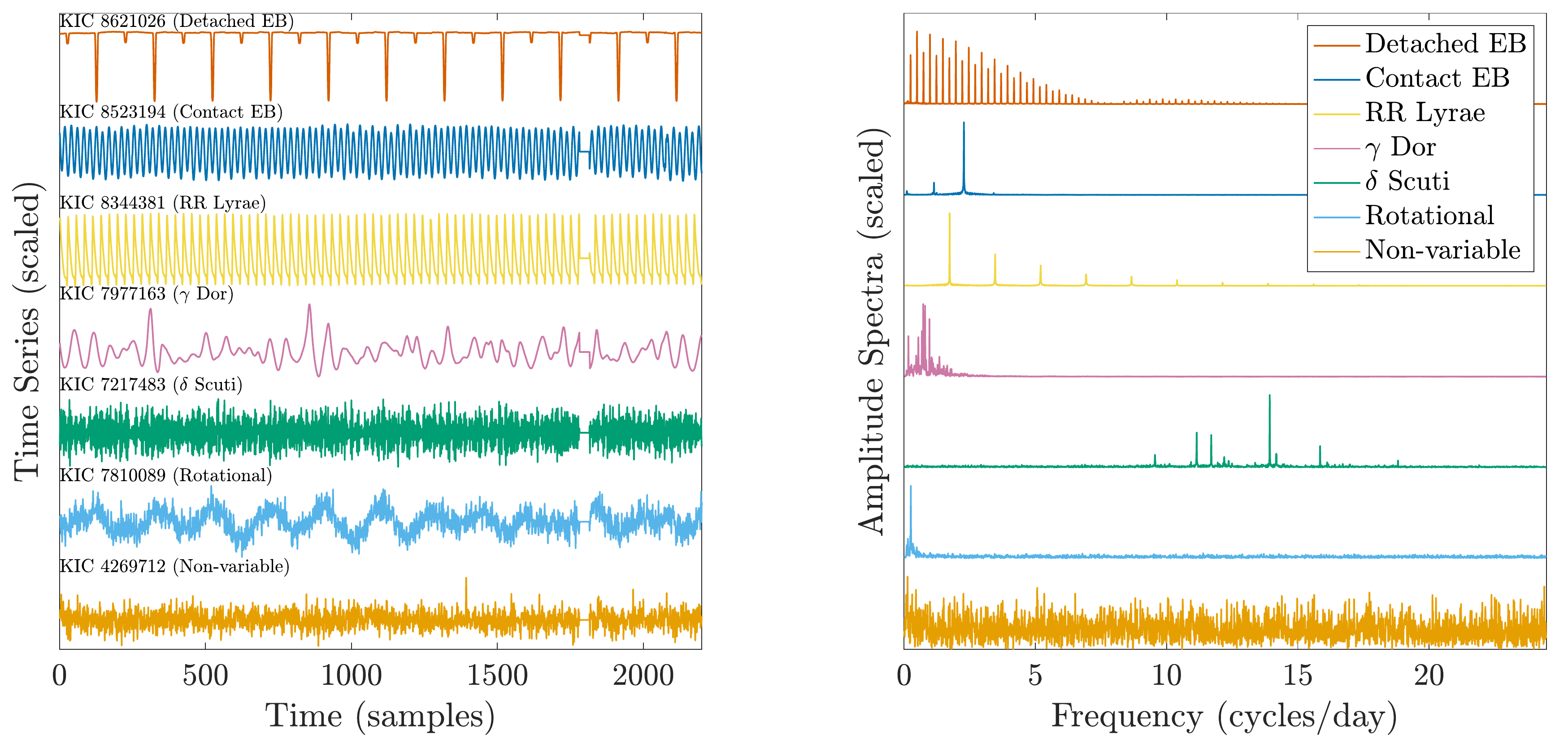}
\caption{Examples of variable stars in the detached binary, contact binary, {\rrl}, {\gdor}, {\dsct}, rotational, and non-variable stellar classes selected in the training set, respectively. The left panel shows the first 2200 samples (at approx. 30 min per sample) of the light curves of each star, or half a {\kep} quarter of data.
The right panel shows the corresponding Fourier transforms up to the Nyquist frequency.
The vertical axes are scaled for ease of viewing.
}
\label{fig:classExample}
\end{figure*}

Selecting an adequate training set to train a feature-based classifier for all 200,000 stars in the {\kep} field is the most challenging and time-consuming aspect of developing a general classification algorithm. Many classes of variable stars are rare, while others remain poorly understood, and still others have not yet been identified. Indeed, new classes are occasionally proposed (\citealt{Debosscher++2011,Bass+Borne2016,pietrukowiczetal2017}). For this reason, a good training set must be representative of a wide range of variable stars to construct a suitably general feature-space representation of {\kep} light curves. Rather than attempting to compile a collection of all known classes of variable and non-variable stars, as attempted with limited success by \cite{Debosscher++2011}, we focused our research on a subset of seven well-studied classes, within the temperature range $6500 \,\text{K} \le T_\mathrm{eff} \le 10,000 \,\text{K} $, as an initial demonstration. The seven chosen classes are $\delta$~Scuti stars, $\gamma$~Doradus stars, RR~Lyrae stars, rotational variables, contact eclipsing binaries, detached eclipsing binaries, and non-variable stars. Typical light curves and power spectra for each class are included in Fig.~\ref{fig:classExample}. These classes are commonly represented in the wider {\kep} data (with the exception of the RR~Lyrae class), are interesting stars that we wish to study in further detail, and are not intermediate or hybrid classes. We excluded hybrid stars to avoid having light curves in the training data that belong to multiple classes.

Two of our seven classes are subclasses of eclipsing binary (EB) systems, which were catalogued in \textit{Kepler} data by \cite{kirketal2016}.
EBs can be classed as one of: detached binaries, where the two stars are far from each other to give highly separated eclipses; contact binaries, where there is no space between the two stellar envelopes, producing almost sinusoidal eclipse patterns; and semi-detached binaries, the intermediate class of the two extremes. In accordance with our decision to exclude hybrid classes, only the detached and contact subclasses have been included in our training set.

Rotational variables are most commonly found among cooler stars, whose star spots present darker patches on the surface that rotate in and out of view, \new{but rotational variability is also seen by \kepler\ across the effective temperature range of our sample \citep{Balona2013, Sikora++2020}.} Typically, the spots have lifetimes not much longer than the rotation period, and they may occur at different latitudes, so the variability is only quasi-periodic \citep{nielsenetal2013,mcquillanetal2014}. The $\alpha^2$\,CVn stars are hotter stars whose strong dipolar magnetic fields
concentrate certain elements into spots. These also rotate with the star, but occur near the magnetic poles and are much longer lived, leading to light curves that do not change rapidly in period, amplitude, or shape \citep{wolff1983}. In our chosen temperature range, examples of both are found.

Three classes of pulsating variable star were included (for a recent review of pulsating stars, see \citealt{Kurtz2022}). RR\,Lyr variables are bright, evolved stars burning helium in their cores. As they traverse the horizontal branch, they cross the instability strip and pulsate periodically with a characteristic phase curve. Their use as standard candles has allowed measurements of the distance to the Galactic centre and to globular clusters
\citep{oort&plaut1975,walker1992}. The two other pulsating star classes, {\gdor} and {\dsct} variables, both comprise A or F-type stars on or near the main sequence, and embody two distinct types of oscillation: g\:modes, or buoyancy-driven modes sensitive to the near-core region of a star; and p\:modes, pressure-driven modes most sensitive to the envelope. {\gdor} stars are multiperiodic g-mode pulsators with periods between approximately 0.3\,d and 3\,d \citep{kayeetal1999}. Despite having periods similar to the RR\,Lyr variables, the multiperiodic {\gdor} stars do not have simple phased curves. There are several hundred in the {\kep} field \citep{lietal2020}, and they have seen substantial recent attention because of their ability to probe internal rotation \citep{vanreethetal2018,ouazzanietal2019}, diffusive mixing \citep{bouabidetal2013}, and core overshooting \citep{mombargetal2019}.

Finally, {\dsct} stars are the most common class of pulsating star at A and F spectral types, with approximately 2000 known in \kep\ data alone \citep{murphyetal2019,Guzik2021}. These stars are p-mode oscillators, and with periods between 18\,min and 8\,hr they are the highest-frequency variables in our sample. For unknown reasons, even in the middle of the {\dsct} instability strip only half of the stars pulsate as {\dsct} stars \citep{murphyetal2019}, hence we include a non-variable class in this work. 
We note that some {\dsct} stars are known to lie outside of the instability strip \citep[e.g.,][]{bowman&kurtz2018}, but our classifications are based only on the {\kep} light curves and not on parameters such as effective temperature. 

\subsection{Preparing the Training Set} \label{ssec:training}

We created a training set across all seven classes by hand-picking 1319 stars from candidate lists according to specific criteria. Stars were restricted to the temperature range $6500 \,\text{K} \le T_\mathrm{eff} \le 10,000 \,\text{K} $ using effective temperatures from \citet{Mathur++2017}.
We examined one quarter of long-cadence {\kep} photometry for each star to prepare the training data. Quarter~9 (Q9) was chosen because it has no prolonged gaps in observation, such as those arising from telescope safe mode events, and no anomalies in data quality. 
We used light curves made with simple aperture photometry (SAP), downloaded from the Kepler Asteroseismic Science Operations Center (KASOC) website (Data Release 25).\footnote{\url{http://kasoc.phys.au.dk/}} The choice to examine a single quarter was made to reduce computation time, but this also precludes the analysis of variability on timescales longer than a typical 90-d quarter. While we certainly recommend the investigation of four-year data in future research focused on a wider range of {\kep} variables, this will not have a great effect on the stars chosen for our investigation. Of the seven classes, only a handful of detached binaries are known to have a period greater than 90 days \citep{kirketal2016}, and we did not include these in the training set.

The training stars were chosen from lists of possible candidates for each of the seven classes by visually inspecting their light curves and Fourier transforms. This laborious process embodies the motivation for automated variable star classification, and was a necessary task to ensure that the training data were accurate and would not mislead the automated feature-selection process. In the following paragraphs, we describe the selection of stars in training sets for each class. Table\:\ref{tab:classes} summarises the class-specific number of stars in the training set. The full list of stars is provided as supplementary material, with a sample shown in Table~\ref{tab:training-set}.

\begin{table}
	\begin{center}
		\caption{Breakdown of stars in the training set.}
		\begin{tabular}{l r}
			\hline
			\textbf{Class} & \textbf{No. Stars} \\
			\hline
			Contact EB & 171 \\
			Detached EB & 83 \\
			{\dsct} & 411 \\
			{\gdor} & 262 \\
			Non-variable & 201 \\
			Rotational & 166 \\
			RR Lyrae & 25 \\
			\hline
			Total & 1319\\
			\hline
		\end{tabular}
		\label{tab:classes}
	\end{center}
\end{table}

\begin{table}
	\begin{center}
		\caption{The training set of 1319 \kepler\ stars. An extract of 14 stars is shown, with the full table provided in the supplementary material.}
		\begin{tabular}{l c}
			\hline
			\textbf{KIC ID} & \textbf{Class} \\
			\hline
			10855535 & Contact EB    \\
			9612468 & Contact EB    \\
			3836439	& Detached EB   \\
            9711751	& Detached EB   \\
            9331207	& {\dsct}       \\
            8376471	& {\dsct}       \\
            4755510	& {\gdor}       \\
            1996456	& {\gdor}       \\
            1864603	& Non-variable  \\
            2156425	& Non-variable  \\
    	    1164109	& Rotational    \\
            1435836	& Rotational    \\
            3733346 & RR Lyrae      \\
            3864443 & RR Lyrae      \\
			\hline
		\end{tabular}
		\label{tab:training-set}
	\end{center}
\end{table}

Eclipsing binary systems were selected from the {\kep} Eclipsing Binary Catalog \citep{kirketal2016}, restricted to periods $<$90\,d and a morphology index of $0 \le c \le 0.5$ (detached binaries) or $0.75 \le c \le 1.0$ (contact binaries), as recommended by \cite{Matijevic++2012}.

Our selection of {\dsct} stars began with 2405 stars manually identified as variable at frequencies above 7\,d$^{-1}$ from a preliminary version of the \citet{murphyetal2019} catalogue. We randomly selected 1000 of these, and further refined this list to remove any stars that were also {\gdor} stars (i.e. {\gdor}/{\dsct} hybrids) by manual inspection. From the same source, we also chose 500 stars that were not {\dsct} pulsators, and removed stars with low-frequency variability to arrive at the 201-star non-variable class.

We selected the {\gdor} sample from the \citet{Debosscher++2011} catalogue by choosing stars with a label confidence of $>$95\% and an effective temperature in the appropriate range. While the \citeauthor{Debosscher++2011} catalogue is known to have errors, this approach was taken due to a lack of an available list of {\gdor} stars exhibiting a broad range of oscillatory behaviours characteristic of the class --- that is, a sample not restricted to neat and well-studied {\gdor} stars from which scientific inference has been made (references in Sec.\,\ref{ssec:classes}). The addition of rigorous manual inspection ensured that the {\gdor} stars included in the final sample were significantly more likely to be correctly classified than in the \citeauthor{Debosscher++2011} catalogue, and that hybrid pulsators were removed.

Unlike the other classes, RR\,Lyr variables are not common in the {\kep} data set. Of the 47 {\kep} RR\,Lyr stars we found in the literature \citep{molnaretlal2018,nemecetal2013,Murphy++2018}, only 25 were observed in Q9. We admitted all 25 of these in the hope that we might discover additional RR Lyrae variables when classifying the remainder of the {\kep} field (we did not).

The rotational variables were selected after trialling a preliminary version of our classifier, trained on the other six classes\new{, on a test sample of \kepler\ stars}. When \new{visually inspecting the} classification results\new{ across the six classes}, we found that rotational variables constituted a considerable fraction of stars (approximately $15-20\%$). From these, we added a list of 166 rotational variables to the training set after a second manual verification.

\subsection{Processing \textit{Kepler} data}
\label{ssec:kep}

Starting with SAP fluxes from Q9 light curves, we processed the data to remove instrumental variability by eliminating long-period trends in the light curve of each star. Such variability can arise from physical drift of the telescope, causing changes in the flux levels falling in the aperture mask, as well as other instrumental effects distinct from stellar variability. Our processing involved division by a smoothed version of each light curve (smoothed using a Savitzky-Golay filter), removal of single-point outliers more than $3\sigma$ from the mean of the smoothed light curve, and converting units to magnitudes.

Any gaps in the data of more than an hour (corresponding to two 29.45-min integrations) were padded with either the mean of the time series for long gaps of four or more integrations, or the mean of the points on either side of smaller gaps. Even in high-quality quarters, long gaps arise from standard telescope operations such as the data downlinks that happen for approximately 24\,hrs twice every quarter, while small gaps may be caused by e.g.\ cosmic ray events. Most machine-learning tools operate as functions of array index rather than explicit functions of time, hence it is imperative that these gaps are filled.

%
%
\section{Feature-based light-curve classification}
\label{sec:classifier}

Having constructed a training set, we next aimed to build a classifier to accurately predict the class of a star from features of its light-curve time series.
Our approach involved four steps:
(i) mapping each light curve to a large feature vector, where each feature is a single, real-valued summary statistic that captures some interpretable property of the light curve;
(ii) learning a classification rule that maps from a reduced subset of extracted features to the class label on a labelled training set;
(iii) evaluating the performance of the learned classification rule on an independent test set; and
(iv) applying this rule to classify the full {\kep} data set.

\subsection{Feature extraction} \label{sec:feature-extraction}

The task of selecting relevant properties of a time series for a given application, like light-curve classification, is commonly a manual one performed by a given researcher (e.g., \citealt{Pashchenko++2017}).
An alternative approach, termed `highly comparative time-series analysis' \citep{Fulcher++2013, Fulcher+Jones2014}, is to include a large and comprehensive candidate set of possible time-series features, and take a data-driven approach to selecting those that are most relevant to the task at hand.
To extract features from a light curve, we used a comprehensive candidate set of over 7000 time-series features from the {\hctsa} software package (v0.96) \citep{Fulcher+Jones2017}.\footnote{\url{https://github.com/benfulcher/hctsa}}
The {\hctsa} feature set encompasses a wide range of time-series analysis methods, from properties of: the distribution of time-series values, linear and nonlinear autocorrelation, entropy and complexity measures, stationarity, time-series model fits, wavelet and Fourier basis-function decompositions, and others \citep{Fulcher++2013}.
This approach allowed us to represent a set of $L$ light curves as an $L \times F$ matrix, where $F$ is the number of features; applying {\hctsa} to our training data set yielded a $1319 \times 7873$ light curve $\times$ feature matrix, where each row is labelled according to one of the seven classes listed in Table~\ref{tab:classes}.
After performing feature extraction, we excluded features that contained special values (\texttt{NaN}, \texttt{Inf}), returned an error, or produced near-constant outputs (within $10 \times$ machine precision) across all 1319 time series, resulting in 6492 features after filtering.
As a preprocessing prior to classification, feature values were normalized to the unit interval using a scaled, outlier-robust sigmoidal transformation \citep{Fulcher++2013}.

\subsection{Training and evaluating a classifier} \label{sec:training-classifiers}

In modern applications of machine learning, complexity is often introduced at the level of the classifier.
In this work we instead focused on selecting from a large candidate set of complex features, but using simple classifiers.
This has the advantage of yielding features that can provide clear scientific interpretation, and follows the approach of \citet{Timmer1993}: ``The crucial problem is not the classificator function (linear or nonlinear), but the selection of well-discriminating features. In addition, the features should contribute to an understanding''.
For classification, we used a Gaussian Mixture Model (GMM) \citep{Mclachlan+Peel2000} on a labelled time series $\times$ feature matrix (described above).
We fitted a single Gaussian component to each of the seven training classes in feature space, combining them with equal prior probabilities to form a seven-component Probability Density Function (PDF).
While all classes are not equally common, equal priors are the simplest choice without knowing the true distribution of variable stars in the \textit{Kepler} field.
Classification was performed by evaluating the (posterior) probability of a star belonging to each class using the trained PDF, and selecting the class with highest probability.
This GMM approach was substantially faster (by factors of approximately 10--100) than alternative algorithms such as nearest-neighbour clustering or support vector machines, but achieved similar classification performance on our training set.

We evaluated classification performance as the average balanced accuracy computed using 10-fold stratified cross-validation \cite{Hastie09}.
Balanced accuracy, $C_\mathrm{bal}$, accounts for class imbalance (the unequal number of observations in each class) in our data set and is defined as:
\begin{equation}
C_{\mathrm{bal}} = \frac{1}{m} \sum_{i=1}^{m}\frac{t_i}{c_i}\,,
\end{equation}
where $m$ is the number of classes, $t_i$ is the number of successfully identified time series in the $i$th class, and $c_i$ is the total number of time series in this class.

\subsection{Feature subset selection}
\label{sec:featureSubsetSelection}

To extract a small number of {\hctsa} features that are most informative of the class labels, we used greedy forward feature selection \citep{Hastie09, Fulcher+Jones2014}.
This simple algorithm iteratively builds a feature set, one feature at a time, with the objective of maximising the balanced classification accuracy, $C_\mathrm{bal}$, at each iteration.
That is, at iteration $k$, the algorithm searches across all individual features for the feature that maximizes $C_\mathrm{bal}$ when used in combination with the features selected in the $k-1$ previous iterations.

{\hctsa} was developed to encompass a comprehensive sample of the interdisciplinary time-series analysis literature, and thus contains groups of features with highly correlated behavior \citep{Fulcher++2013, Henderson2021:EmpiricalEvaluationTimeSeries}.
When multiple features exhibit similar classification performance, we implemented a simple heuristic constraint to favour the selection of faster-to-compute features: at each iteration, of the features with an accuracy within a margin of 1\% of the best-performing feature, the feature with the fastest computation time was selected.
The iterative procedure was terminated when the improvement in training-set $C_\mathrm{bal}$ from adding another feature dropped below 1\%.
Note that applying this algorithm to the full training data set has the potential to overfit, since the selection step at each iteration (despite using cross-validation for each feature) uses the training set itself to select the best-performing feature.
Accordingly, we evaluate the performance of our reduced feature set on an independent test set in Section~\ref{sec:results}.

%
%

\section{Results \& Discussion}
\label{sec:results}

\subsection{Representing light curves in a high-dimensional feature space}

\begin{figure}
    \centering
	\includegraphics[scale=0.47]{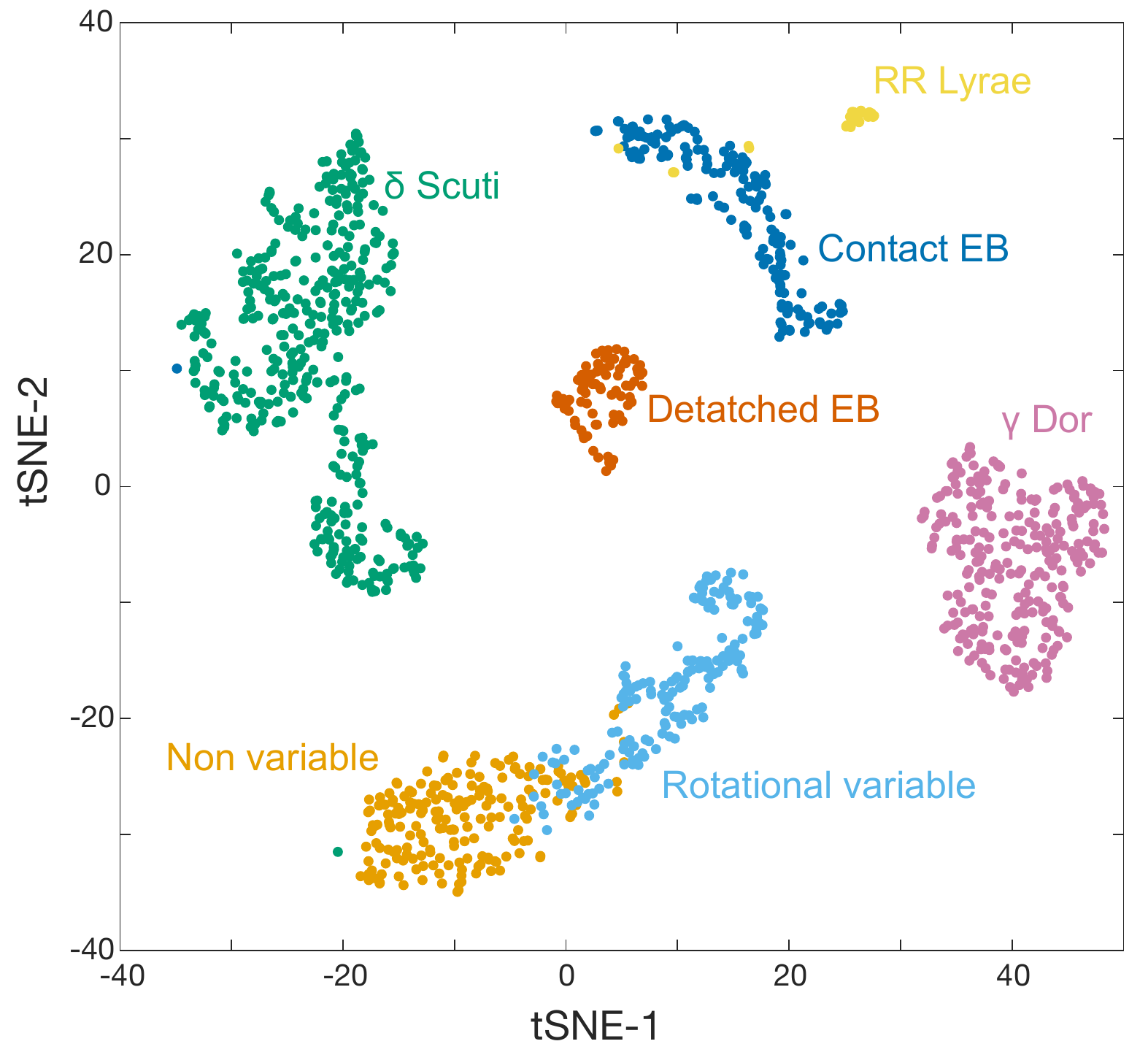}
	\caption{
	A two-dimensional $t$-SNE projection of our {\kep} training set of 1319 stars in the 6492-dimensional {\hctsa} feature space, where each light curve is coloured by its class label.
	Most stars form clear clusters that match their class identity, indicating that the {\hctsa} features provide a useful space in which to represent \textit{Kepler} light curves.
	}
	\label{fig:tSNE_training}
\end{figure}

We first investigated the structure of the seven labelled classes of 1319 {\kep} stars in the 6492-dimensional {\hctsa} feature space.
We found that the {\hctsa} feature space is able to capture characteristic properties of the seven labelled classes of stars, obtaining a high mean 10-fold cross-validated balanced accuracy of 95.9\% (using a linear support vector machine \citep{Hastie09}, compared with a chance rate for seven classes of 14.3\%).
This indicates that each type of star displays distinctive dynamics in ways that can be detected by the features included in {\hctsa}.
To better understand the structure of light curves in the high-dimensional {\hctsa} feature space, we inspected a two-dimensional $t$-SNE visualization ($t$-distributed stochastic neighbour embedding; \citealt{vanderMaaten2008:VisualizingDataUsing}).
The result is shown in Fig.~\ref{fig:tSNE_training}, where each point is a light curve, and light curves with similar features tend to be positioned closely in the space.
While $t$-SNE is an unsupervised technique (Fig.~\ref{fig:tSNE_training} was constructed without knowledge of the class labels), stars are meaningfully organized according to their labelled class, with most stars clustering with other stars of the same type.
Consistent with the high classification results reported above, this indicates that the {\hctsa} feature space captures distinctive dynamical properties of the light curves corresponding to the seven different types of stars.
The plot also reveals scientifically meaningful structure between classes, such as the continuum from non-variable (light orange) stars to rotational-variable (light blue) stars.
\new{We also see a small overlap between \rrl\ stars and contact EBs, which reflects the similar morphologies of their light curves (see Fig.~\ref{fig:classExample}).}


\subsection{Representing light curves in a reduced feature space}

\begin{figure}
\includegraphics[width=\linewidth]{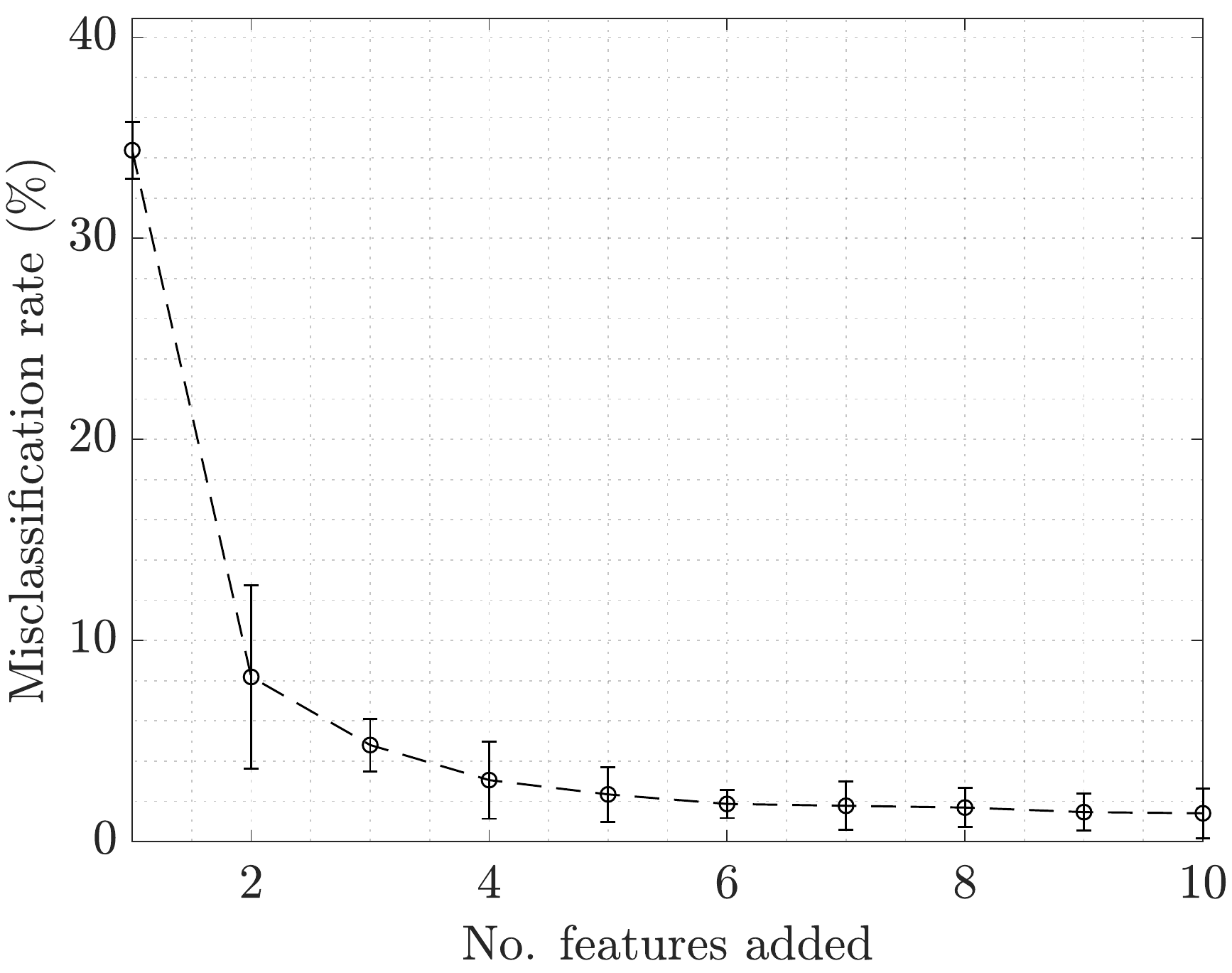}
\caption{
Classification performance as a function of the number of time-series features.
Balanced misclassification rate on the training set (using a GMM classifier) is plotted as a function of selected features, shown as the mean and standard deviation across 10-fold cross validation.
}
\label{fig:misclass}
\end{figure}

The results above demonstrate that time-series properties in {\hctsa} can capture differences in light-curve dynamics between different types of stars.
But which types of individual time-series features are most informative of these differences?
To address this question, we aimed to construct a reduced set of {\hctsa} features that display strong classification performance using greedy forward selection (see Sec.~\ref{sec:featureSubsetSelection} for details).
The cross-validated balanced misclassification rate on the training set is shown as a function of the number of selected features in Fig.~\ref{fig:misclass}.
This plot reveals that strong in-sample classification performance can be obtained with a relatively small set of well-chosen time-series features, e.g., a balanced accuracy of 95.2\% with just three features.
According to our termination criterion---when an additional feature provides $<1$\% marginal improvement in balanced accuracy---we obtained an informative five-dimensional feature space in which to represent {\kep} light curves.

\begin{figure*}
	\begin{center}
		\hfill
		{\includegraphics[width=0.4\linewidth]{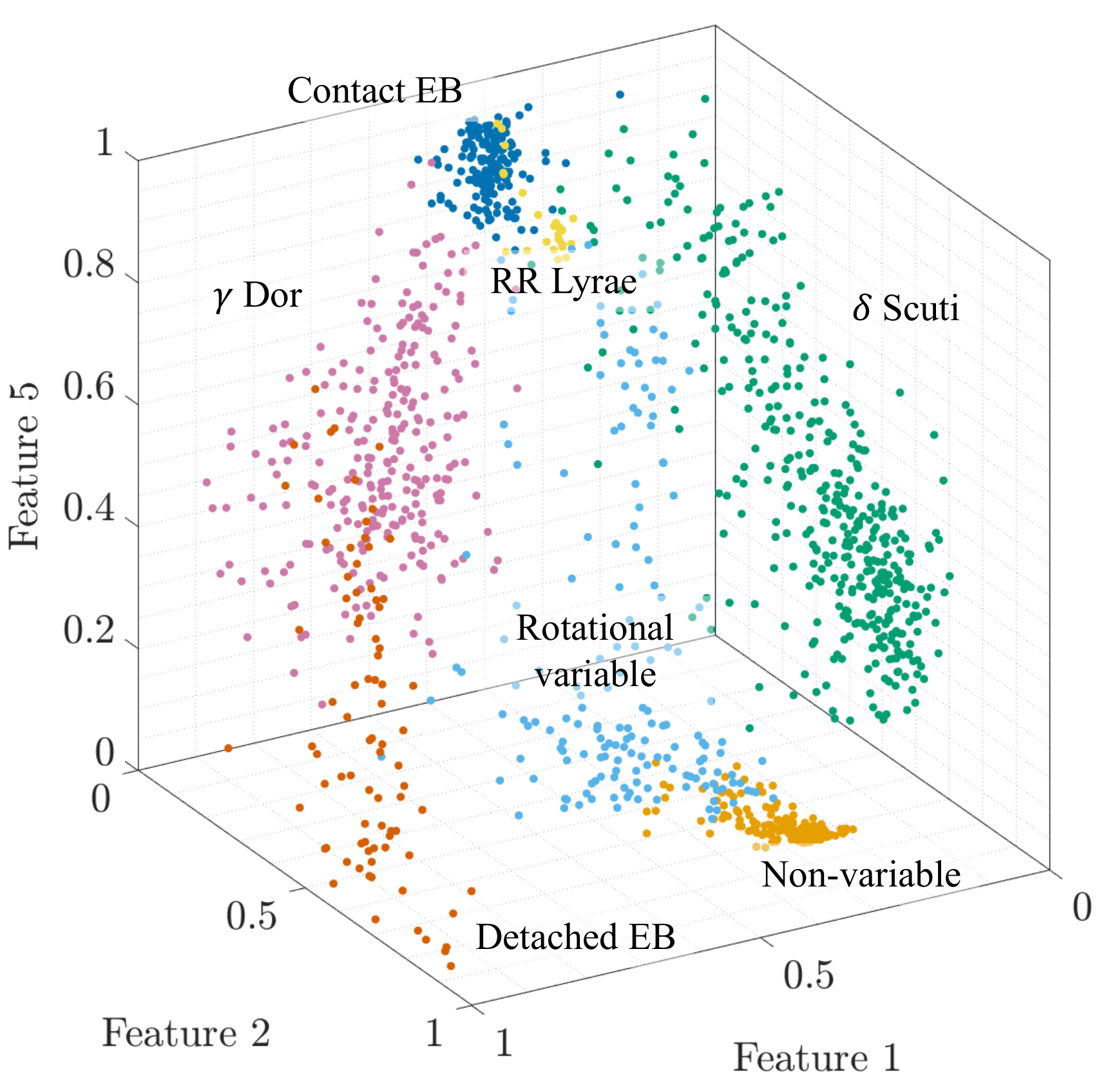}} 
		\hfill
	    {\includegraphics[width=0.47\linewidth]{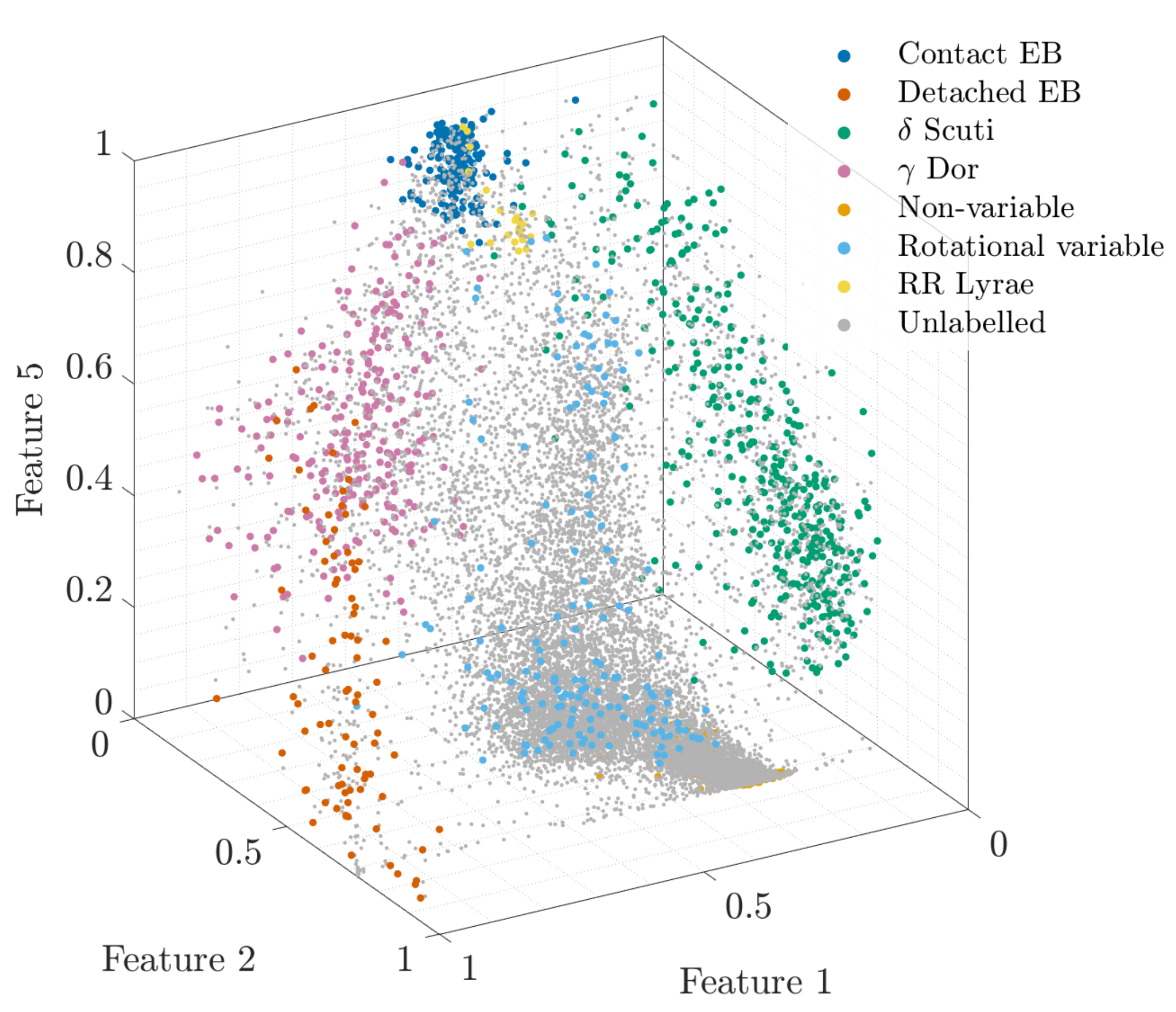}}
		\hfill
	\end{center}
	\caption{
	Separation of different classes of stars in the (normalized) space of three time-series features selected from {\hctsa} using greedy forward-feature selection.
	The two figures show (left) the training set, and (right) the training set alongside unlabelled {\kepler} data with $6500 \,\text{K} \le T_\mathrm{eff} \le 10,000 \,\text{K} $.
	The three features correspond to \texttt{AC\_nl\_001}, \texttt{MF\_steps\_ahead\_ar\_best\_6\_mabserr\_5}, and \texttt{SP\_Summaries\_welch\_rect\_peakPower\_5}, as described in detail in Sec.~\ref{sec:reduced-features}.
	}
	\label{fig:featureSpace}
\end{figure*}

To visualize how stars are organized in the reduced feature space, we plotted the training set in the space corresponding to three of the selected features in Fig.~\ref{fig:featureSpace} (left).
Despite a dramatic dimensionality reduction of each time series---from the 4767 data points in a typical Q9 time series to just three extracted summary statistics---the space meaningfully organizes all seven training classes in this low-dimensional feature representation, with each occupying a characteristic region of the space. Much like the $t$-SNE construction in Fig.~\ref{fig:tSNE_training}, the relative positions of each class are consistent with what we would intuitively expect from their light curves and power spectra in Fig.~\ref{fig:classExample}. For example: detached binaries are highly separated from the other classes, as their light curves are the most distinct; non-variable stars blend with rotational variables when the rotations are weak and difficult to distinguish by eye, such that the light curves are almost non-varying; the {\gdor} and {\dsct} stars lie on opposite sides of the space, reflecting their contrasting low- and high-frequency pulsations; and the contact binaries are close to the {\gdor} and RR Lyrae clusters, which all characteristically exhibit regular low-frequency variability.


\subsubsection{The reduced feature set} \label{sec:reduced-features}

We have demonstrated the usefulness of representing {\kep} light curves in an low-dimensional feature space, but what types of properties are these features measuring, and what can that tell us about how light-curve dynamics differ between the seven classes of stars?
In this section, we explain the five features in order of their selection by our greedy forward selection algorithm.
Noting the small marginal improvements in accuracy after approximately three features (shown in Fig.~\ref{fig:misclass}), we focus in particular on these features.
In the following discussion, note that the time series were converted to magnitudes, so that positive excursions correspond to decreases in stellar flux, and vice versa.

\begin{figure}
\includegraphics[width=0.95\linewidth]{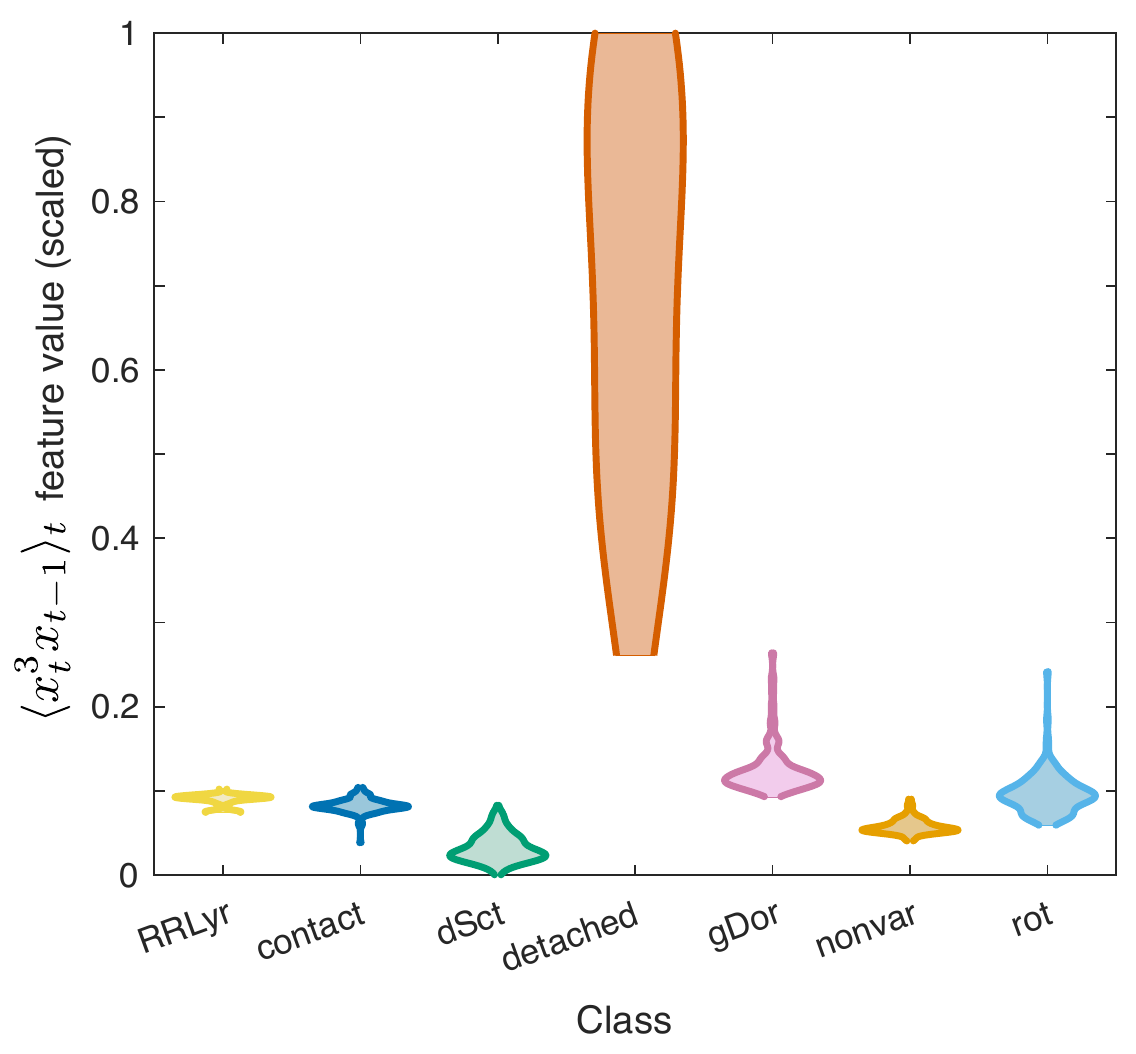}
\caption{
Values for Feature 1, which computes $\langle x_t^3 x_{t-1} \rangle_t$ (see Sec.~\ref{sec:reduced-features}).
The violin plots show the normalised output of \texttt{AC\_nl\_001} across the seven classes of stars in the training set.
Sigmoidal normalization scaled to the unit interval (see methods) was used to aid visualisation of the large range of raw values of this feature.
}
\label{fig:feature1_violin}
\end{figure}

The first selected feature (labelled \verb|AC_nl_001| in {\hctsa}), is a nonlinear autocorrelation statistic that computes the time-average, $\langle x_t^3 x_{t-1} \rangle_t$, of the $z$-scored time series $x_t$, with a time-lag of 1 sample (approximately 30 minutes in the time domain).
Similar to a lag-1 autocorrelation, $\langle x_t x_{t-1} \rangle_t$, it gives high values to highly autocorrelated light curves, but the modification ($x_t^3$) accentuates large deviations from the mean.
The distribution of this feature's (sigmoid-normalized) values across the seven classes of stars is shown in Fig.~\ref{fig:feature1_violin}.
Detached binaries have the largest values of this statistic, driven by large positive excursions from the mean (since the time series are in magnitudes).
Autocorrelation arising from slower periodic patterns, as in {\gdor}, rotational stars, {\rrl} and contact binaries, lead to moderate positive values of \verb|AC_nl_001|, while the non-variable stars have low values (raw values near zero).
The high-frequency oscillations seen in some {\dsct} stars \citep[e.g.,][]{Balona++2019,beddingetal2020} resulted in negative values of \verb|AC_nl_001| (the lowest normalized values).

\begin{figure}
\includegraphics[width=0.95\linewidth]{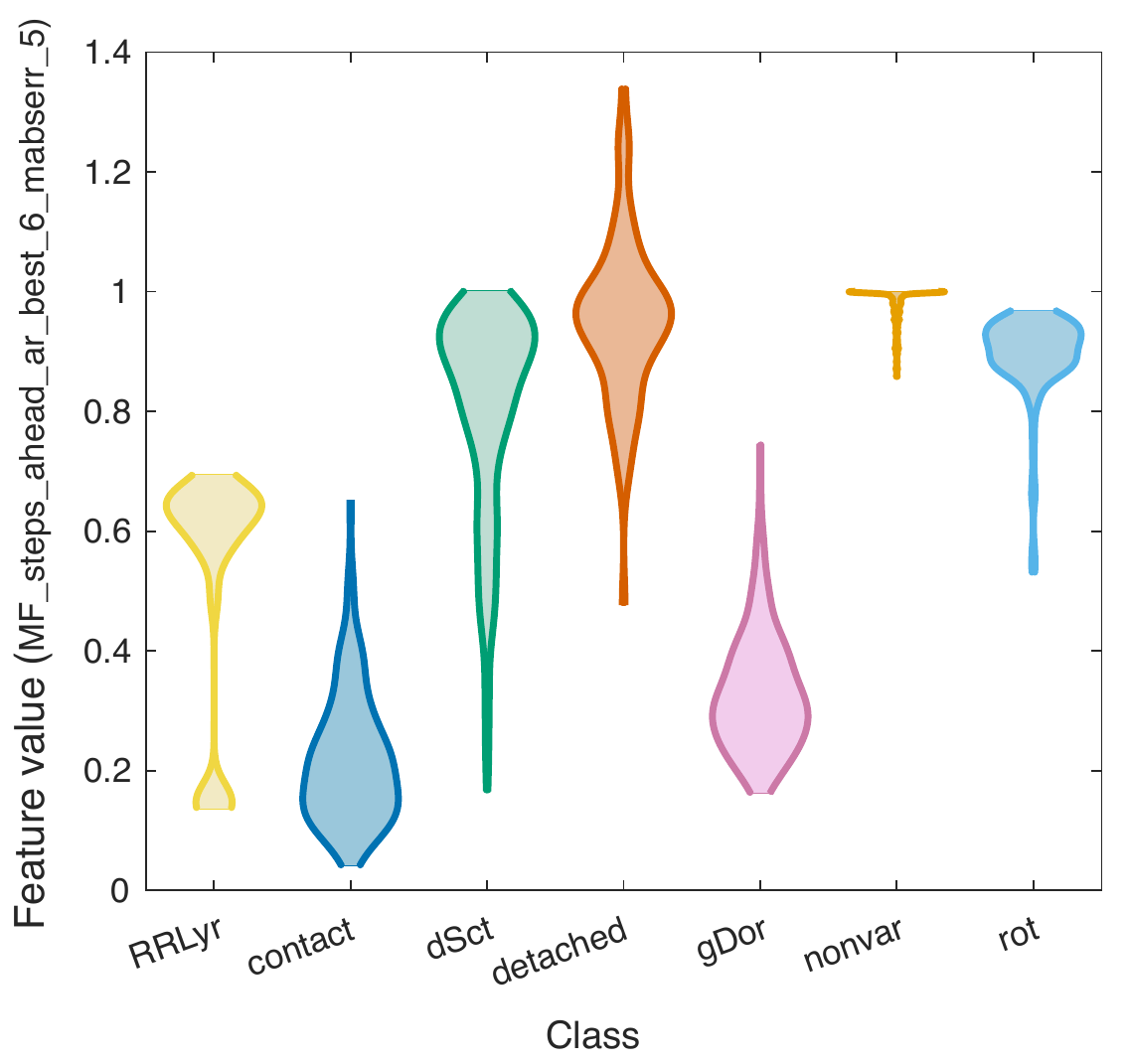}
\caption{
Distribution of Feature 2 values by class (see Sec.~\ref{sec:reduced-features}), which measures how predictable the time series is using a linear autoregressive (AR) model; high values (near 1) are given to light curves for which the AR model performs worse than simple benchmarks, whereas values near 0 are given when the AR model strongly outperforms the benchmarks.
Violin plots are shown for the distribution of this feature across the seven classes of stars.
}
\label{fig:feature2_violin}
\end{figure}

Feature~2 (labelled \verb|MF_steps_ahead_ar_best_6_mabserr_5| in {\hctsa}) uses a linear autoregressive (AR) model to measure how predictable a time series is.
This statistic captures how well an AR model (of optimal order, selected in the range 1--10 using Schwartz's Bayesian Criterion) can predict 5 time steps ahead in the time series. This is measured relative to simple benchmark forecasting methods (including simple mean forecasts and a constant global-mean forecast), calculated as the mean absolute error.
The distribution of feature values across the seven classes of stars is shown in Fig.~\ref{fig:feature2_violin}.
Values near zero indicate strong prediction performance of the AR model relative to simple benchmarks, while values greater than 1 indicate relatively inferior model performance.
We see high values for the non-variable stars, detached binaries, rotational stars, and most of the {\dsct} stars, with {\rrl} stars displaying intermediate values
\new{(a few \rrl\ stars with highly symmetric light curves have low values).}
The {\gdor} and contact binary light curves exhibit a strong linear correlation structure that allowed the AR models to make strong forecasts of these time series, yielding low values for this statistic.

Feature~3, labelled \verb|CO_trev_3_num| in {\hctsa}, evaluates the following time-average: $\langle (x_t - x_{t-3})^3 \rangle_t$.
This statistic, using a time-lag $\tau = 3$, can be thought of as capturing asymmetry in the size of increases ($x_t - x_{t-3} > 0$) versus decreases ($x_t - x_{t-3} < 0$).
For example, time series with sudden increases but gradual decreases (at a time-lag of 3 samples) will have large values of this feature.
{\rrl} are distinguished by negative values of \verb|CO_trev_3_num|, due to the characteristic asymmetry in the shapes of their light curves \citep[e.g.,][]{Catelan+Smith2015}.

Feature~4, labelled \verb|ST_LocalExtrema_n100_medianmax| in {\hctsa}, captures how positive outliers are distributed through the time series.
Operating on the $z$-scored time series, this algorithm computes the maximum value in each of 100 overlapping windows (each containing 47 samples corresponding to approximately 23 hours), and outputs the median of these local maxima.
For time series with relatively infrequent large positive excursions (like the light curves from many detached binaries, recalling that the calculations are done with magnitudes), most windows will have very low maxima, and thus the median of the maxima will be a low value.
But for time series with maxima spaced more evenly throughout time, like most non-variable and {\dsct} stars, high values are obtained for this statistic.

Feature~5, labelled \verb|SP_Summaries_welch_rect_peakPower_5| in {\hctsa}, uses Welch's method and a rectangular window to estimate the power spectrum and returns the proportion of power captured by the five most prominent identified peaks.
Broadly, this feature gives high values to time series that are well-captured by a relatively small number of dominant frequencies.
The lowest values for this feature were found for non-variable stars and rotational variables, while high values were obtained for contact binaries and {\rrl} stars.

\subsubsection{Evaluation on a test data set}
\label{sec:test-set}

Having computed an informative low-dimensional space in which to represent {\kep} light curves, we investigated its effectiveness in classifying variable stars outside our training set. We manually compiled a test set of 515 stars in the {\kep} field belonging to classes of variable stars in our training set, and with $6500 \,\text{K} \le T_\mathrm{eff} \le 10,000 \,\text{K}$. The full list of test stars is provided as supplementary material, with a sample shown in Table ~\ref{tab:test-set}.

\begin{table}
	\begin{center}
		\caption{The test set of 515 \kepler\ stars. An extract of 12 stars is shown, with the full table provided in the supplementary material.}
		\begin{tabular}{l c}
			\hline
			\textbf{KIC ID} & \textbf{Class} \\
			\hline
			8282730 & Contact EB    \\
			6957185 & Contact EB    \\
			8953296	& Detached EB   \\
            5090690	& Detached EB   \\
            8585472	& {\dsct}       \\
            3648131	& {\dsct}       \\
            6041803	& {\gdor}       \\
            8739181	& {\gdor}       \\
            5616145	& Non-variable  \\
            8153411	& Non-variable  \\
    	    3847563	& Rotational    \\
            3967219	& Rotational    \\
			\hline
		\end{tabular}
		\label{tab:test-set}
	\end{center}
\end{table}

\begin{figure}
\includegraphics[width=\linewidth]{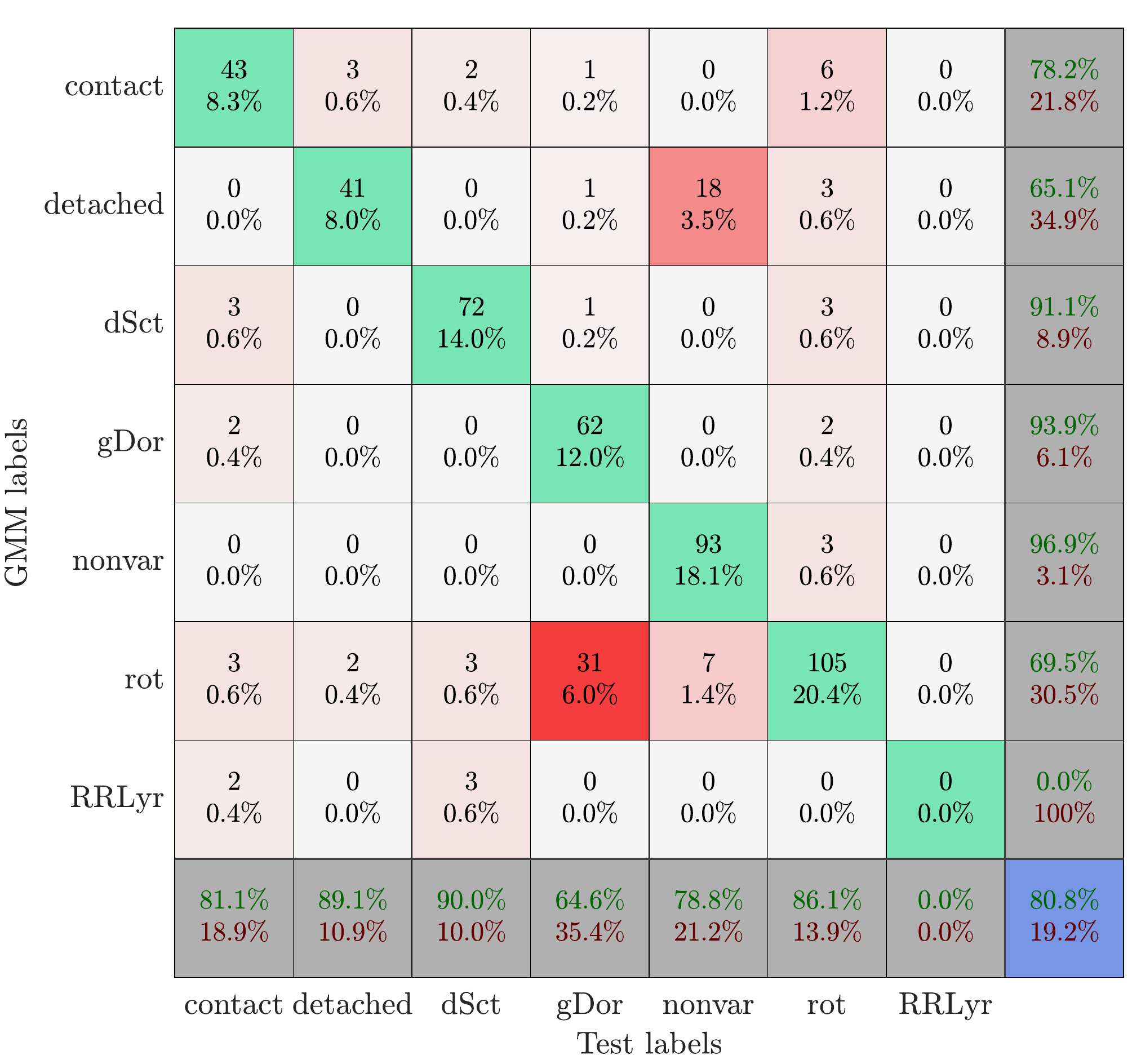}
\caption{Confusion matrix summarising GMM classification performance. The GMM and test labels are the classifier predictions and manually-assigned truth labels (respectively) for stars in our test set. Summaries in grey on the right of the matrix correspond to the (unbalanced) percentage of correct predictions, while summaries at the bottom are the (unbalanced) percentage of each class that was correctly classified. The raw classification accuracy is shown in blue (balanced accuracy 81.6\%). There were no RR~Lyrae stars in our test set.}
\label{fig:confusion}
\end{figure}

To evaluate classification performance on the test set in the trained five-dimensional feature space, we constructed a GMM consisting of seven Gaussian components, one fitted to each class in our training set (with uniform priors), and used it to classify each of the test stars.
Figure~\ref{fig:confusion} summarises our results on the test set as a confusion matrix. 

\new{The confusion matrix can be interpreted as follows. Labels on each row were assigned by the GMM classifier, while labels on each column correspond to manually-assigned labels from our test set. Each cell $(i,j)$ of the confusion matrix shows the number of stars (and percentage of all stars considered) that were classified as category $i$ by the GMM, and as category $j$ in our test set. For example, there were 18 stars classified as detached binaries by the GMM but labelled as non-variable in the test set. Diagonal elements of the matrix (in green) correspond to correctly classified stars. Summaries in grey on the right of the matrix correspond to the (unbalanced) percentage of correct predictions, while summaries at the bottom are the percentage of each class that was correctly classified. The raw classification accuracy is shown in blue in the bottom-right corner.}

Our classifier achieved a balanced accuracy of 81.6\% on the test set and performed well on all classes, with two understandable exceptions \new{highlighted in Figure \ref{fig:confusion}}:
\begin{itemize}
    \item Non-variable stars are commonly misclassified as detached binaries (18 misclassifications). Most have sharp transitions in their light curves at the beginning or end of the quarter, or just before or after the {\kep} telescope paused observation for data transmission. These transitions appear as sharp peaks or troughs, and are represented similarly to eclipses in our feature space.
    
    \item {\gdor} stars are commonly misclassified as rotational variables. Both classes have low-frequency variations (e.g. \citealt{glietal2019b}) and even for an expert eye, it can be difficult to resolve {\gdor} oscillations from a single quarter of \textit{Kepler} data. The behaviour may therefore look similar to rotation in our feature space.
\end{itemize}
Apart from these exceptions, our approach yielded high overall classification accuracies despite relying on very simple methods (greedy forward feature selection and GMM classification), demonstrating the usefulness of the comprehensive {\hctsa} feature space in highlighting high-performing interpretable features for a given problem.
We expect that repeating the feature selection and classification procedures with more sophisticated \new{algorithms, while still working with a rich set of interpretable features,} would further improve the accuracy reported here using simple methods.
However, as discussed in Section~\ref{sec:classify-kepler}, our method is already a useful tool for classifying and searching large data sets.

\subsection{Classifying the \textit{Kepler} field}
\label{sec:classify-kepler}

In this section, we use the low-dimensional feature space learned from the training set and validated on the test set to classify variable stars across the entire {\kep} field. Our full classification catalogue is provided as supplementary material (Table~\ref{tab:posteriors}).

\subsubsection{Classifying unlabelled stars}

We computed the 5-dimensional feature-space representation of all 12,088 stars with Q9 data in the \textit{Kepler} field and $6500 \,\text{K} \le T_\mathrm{eff} \le 10,000 \,\text{K}$ (excluding our training set of 1319 stars). These are plotted in the right panel of Fig.~\ref{fig:featureSpace} as unlabelled stars (grey). Each feature vector was normalized using the same scaled robust sigmoid transformation (including its coefficients) as used on the training set, preserving the structure of our normalized feature space. The grey unlabelled stars are clearly clustered around the coloured training groups, with the majority of stars residing near the non-variable cluster. This clustering occurs naturally because of our choice of feature space. Intuitively, we might expect that (i)~unlabelled stars near each training cluster belong to that respective class; (ii)~stars midway between two groups are hybrids of both classes; and (iii)~stars far from any group are new classes of variable stars unaccounted for in our training set.
We have already verified the first of these hypotheses by applying our GMM classifier to the test set with reasonably high accuracy in Section~\ref{sec:test-set}. We leave investigation of the remaining two claims as future work.

We evaluated our trained GMM classifier on all 12,088 unlabelled stars to generate a catalogue of posterior probabilities, giving a predicted classification for each star. The first ten lines of this catalogue are shown in Table~\ref{tab:posteriors}, with the full catalogue provided as supplementary material. For each star in the catalogue, its classification is the class with maximum posterior probability. The catalogue is intended as a useful tool in searching for candidate variable stars of interest. We note that this catalogue and our broader methodology have already proven useful in identifying new {\gdor} stars \citep{glietal2019b} and {\dsct} stars \citep{murphyetal2020a} in the {\kep} field. We provide suggestions for searching our catalogue in Section~\ref{sec:pdf}.

\begin{table*}
	\begin{center}
	\caption{Extract of GMM posterior class probabilities and probability density $p(x)$ for 12,088 unlabelled stars in the \textit{Kepler} field. The first 10 lines are shown, with the full table provided in the supplementary material.}
	\label{tab:posteriors}
	\begin{tabular}{@{}lcccccccc@{}}
	\hline
	\textbf{KIC ID} & \textbf{Contact EB} & \textbf{Detached EB} & \textbf{$\delta$ Scuti} & \textbf{$\gamma$ Dor} & \textbf{Non-variable} & \textbf{Rotational variable} & \textbf{RR Lyrae} & \textbf{$p(x)$} \\
	\hline
    757280  & 0.00  & 0.00  & 0.00  & 0.00  & 0.00  & 1.00  & 0.00  & 0.73      \\
    892667  & 0.00  & 0.00  & 0.00  & 0.00  & 0.00  & 1.00  & 0.00  & 9.12      \\
    892828  & 0.00  & 0.00  & 0.00  & 0.00  & 0.93  & 0.07  & 0.00  & 307.19    \\
    893234  & 0.00  & 0.00  & 0.00  & 0.00  & 0.01  & 0.99  & 0.00  & 4.12      \\
    893944  & 0.00  & 0.00  & 0.00  & 0.00  & 1.00  & 0.00  & 0.00  & 3802.55   \\
    1026133 & 0.00  & 0.00  & 0.00  & 0.00  & 0.00  & 1.00  & 0.00  & 1.02      \\
    1026255 & 0.00  & 0.00  & 0.00  & 0.00  & 0.51  & 0.49  & 0.00  & 0.71      \\
    1026475 & 0.00  & 0.00  & 0.00  & 0.00  & 0.00  & 1.00  & 0.00  & 9.54      \\
    1026861 & 0.00  & 0.00  & 0.00  & 1.00  & 0.00  & 0.00  & 0.00  & 0.44      \\
	\hline
	\end{tabular}
	\end{center}
\end{table*}

\subsubsection{Using probability density as a confidence heuristic}
\label{sec:pdf}

Close examination of Fig.~\ref{fig:featureSpace} reveals that many unlabelled stars lie in areas between the training set clusters, far from where the GMM classifier was trained. We may therefore ask: how does our classification accuracy improve if we restrict the test set to stars ``near'' the training distributions? We define $p(x)$ as the probability density of a star represented by feature vector $x$, where the probability density function is the 7-class GMM used for classification. Stars close (in feature space) to the centre of the multivariate Gaussians will have large probability densities, $p(x)$, while those far from the class centroids will have low $p(x)$. In this sense, we can use $p(x)$ as a heuristic measure of how likely a star is to belong to any of the training classes. $p(x)$ is provided in the final column of Table \ref{tab:posteriors}.

\begin{figure*}
	\begin{center}
		\hfill
		{\includegraphics[width=0.45\linewidth]{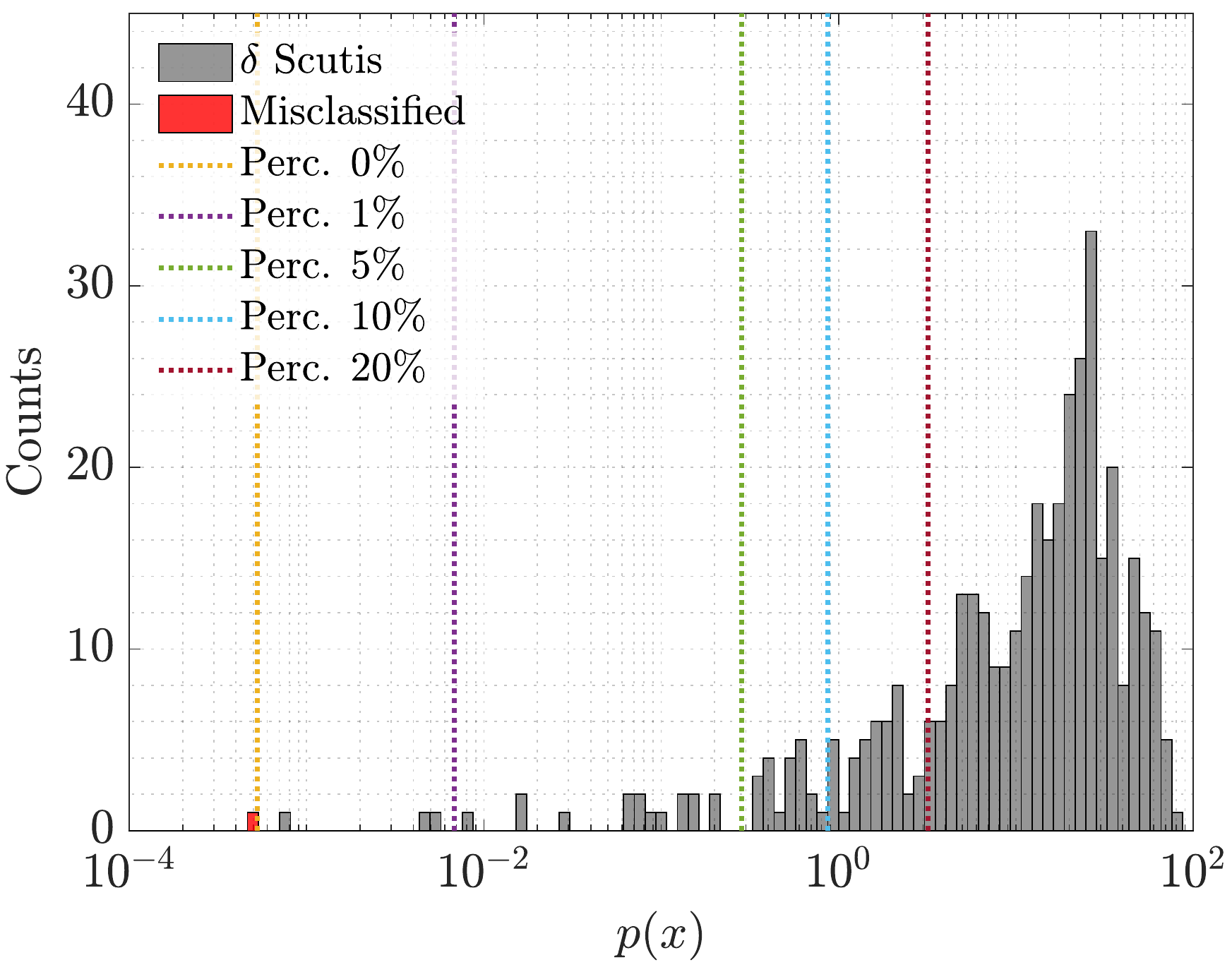}}
		\hfill
	    {\includegraphics[width=0.45\linewidth]{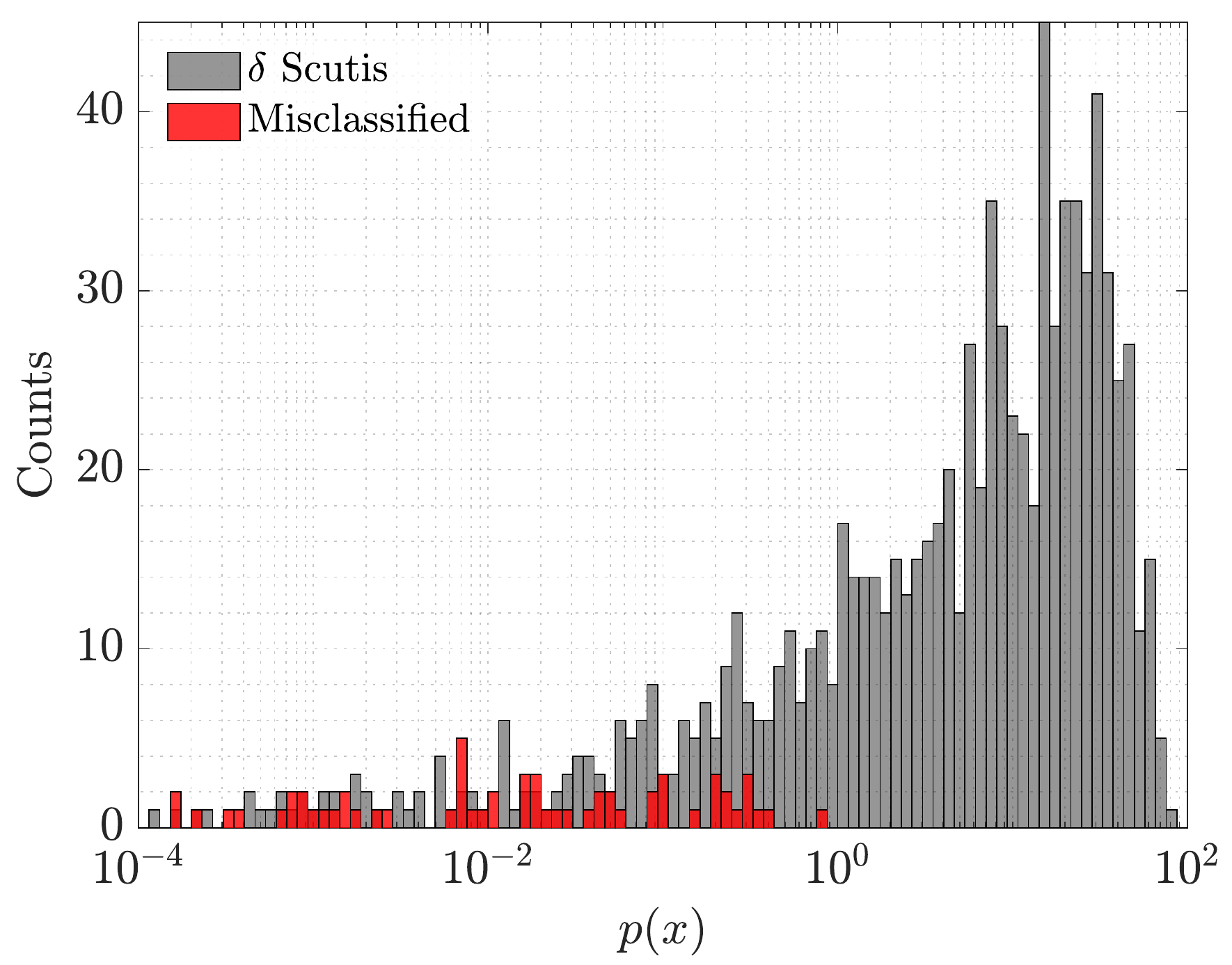}}
		\hfill
	\end{center}
	\caption{
	Histograms of GMM probability density $p(x)$ for all stars classified as $\delta$ Scutis in (left) the training set, and (right) the remainder of the \textit{Kepler} field with $6500$\,K $\le T_\mathrm{eff} \le 10,000$\,K. Lines in the training distribution indicate $p(x)$ percentile cut-offs. For example, 90\% of stars classified as $\delta$ Scuti in the training set lie above the blue line. Stars in the right panel are classified according to \citet{murphyetal2019}.
	}
    \label{fig:dsct-px}
\end{figure*}

As an example, the left and right panels of Fig.~\ref{fig:dsct-px} show the distributions of $p(x)$ for all {\dsct} stars in the training set and the rest of the \textit{Kepler} field (in our temperature range of interest), respectively. Classifications in the right panel of Fig.~\ref{fig:dsct-px} are from the (manually-compiled) \cite{murphyetal2019} catalogue. As anticipated, misclassification is far more common at low densities. Interestingly, the distribution in Fig.~\ref{fig:dsct-px} shows that above $p(x) \approx 1$, all predictions of {\dsct} stars are correct. We would intuitively expect similar $p(x)$ cutoffs for the other classes, above which we have high confidence in the GMM predictions. However, defining exact cut-offs is impossible without a full labelled catalogue of the \textit{Kepler} field. Instead, we can define values for $p(x)$ representing regions of increasing proximity to our trained distribution. The vertical lines in the left panel of Fig.~\ref{fig:dsct-px} show $p(x)$ percentile cut-offs, above which a certain percentage of the training data fall. For example, only the 90\% ``closest'' {\dsct} training stars to the {\dsct} centroid (in terms of probability density) lie above the blue line in Fig.~\ref{fig:dsct-px}.

\begin{figure}
\includegraphics[width=\linewidth]{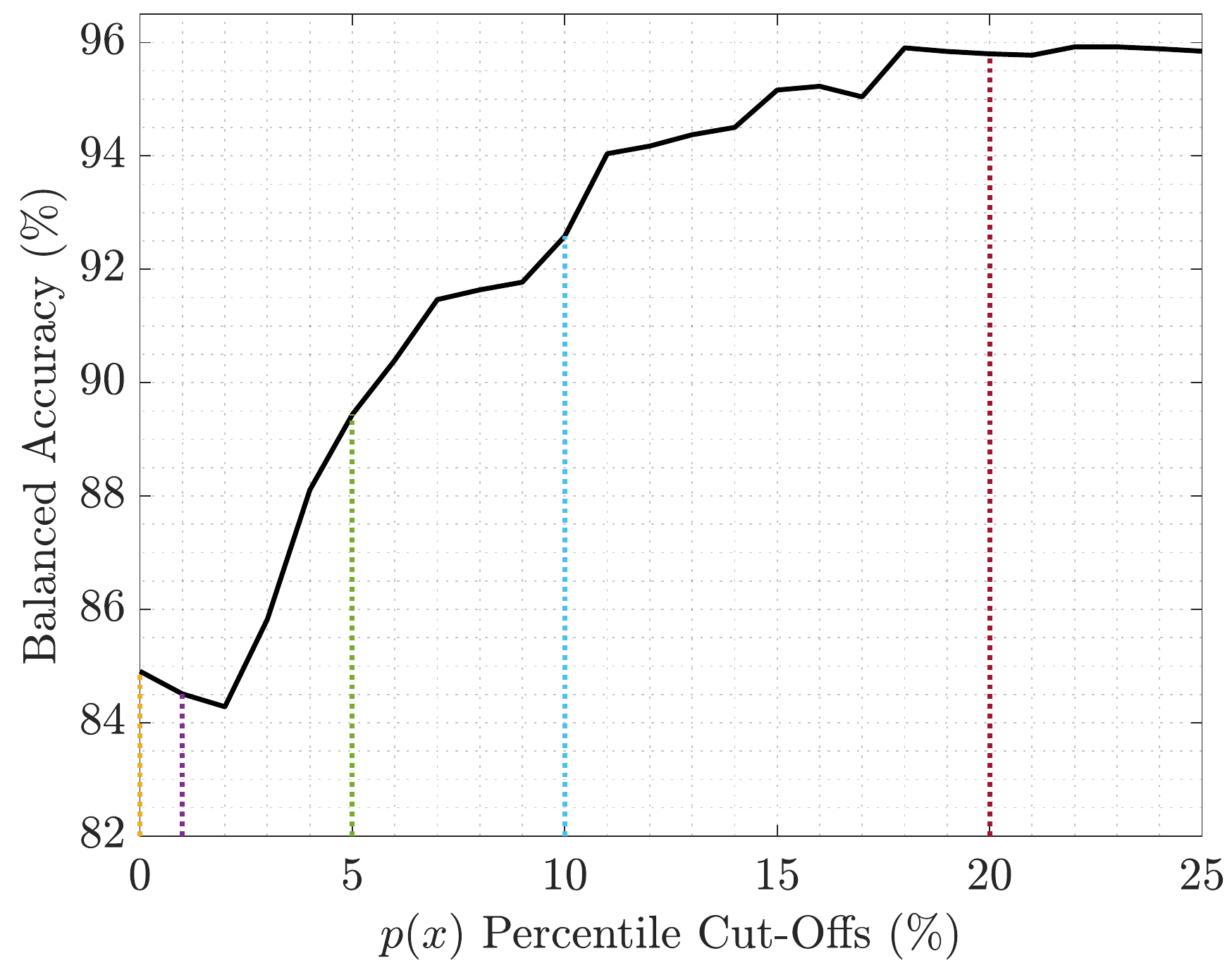}
\caption{Balanced classification accuracy as a function of $p(x)$ percentile cut-offs. Applying the classifier to stars close to regions of feature-space that we trained on significantly improves the overall accuracy. Dotted lines correspond to the same percentile cut-offs overlaid in Fig.~\ref{fig:dsct-px}.}
\label{fig:balacc_px}
\end{figure}

The results above, for {\dsct} stars, suggests that our predictions are more accurate in higher-confidence areas of the feature space, corresponding to areas with higher modelled density for the training set.
To test whether this holds more generally, we computed the balanced classification accuracy (across all classes) on the test set for a range of $p(x)$ percentile thresholds.
As shown in Fig.~\ref{fig:balacc_px}, we find that accuracy improves with more stringent restrictions on $p(x)$, demonstrating the usefulness of $p(x)$ as a proxy for prediction confidence. Even small restrictions in $p(x)$, such as the $95^\text{th}$ percentile cut-off (green line), improve the classification performance on our test set to approximately 90\% accuracy. This is an example of a useful way to search our catalogue and obtain a list of confidently-classified variable stars for further analysis --- as $p(x)$ increases for each class, so too does the confidence of our predictions. We once again stress that such intuitive search criteria are a direct consequence of our choice of feature space and simple classification algorithm. One could achieve even more accurate results with more sophisticated approaches, but this may come at the expense of interpretability of our low-dimensional feature space.

\subsection{Comparison with Audenaert et al.\ (2021)}
\label{sec:audenaert}

When our paper was in the final stages of preparation, a new classification of \kepler\ light curves was published by \citet[][hereafter Aud21]{Audenaert++2021}.  Their work was done as part of efforts to design an automated classification algorithm for the {\tess} mission.  Given the complementary nature of Aud21 and our own study, especially given that both were based on Q9 data, it is worthwhile to carry out a brief comparison.  We should keep in mind that the emphasis in Aud21 was on providing a high-performance classification pipeline from existing methods, whereas ours involved designing an interpretable classifier from a rich library of time-series features.

\begin{figure}
	\begin{center}
	    {\includegraphics[width=\linewidth]{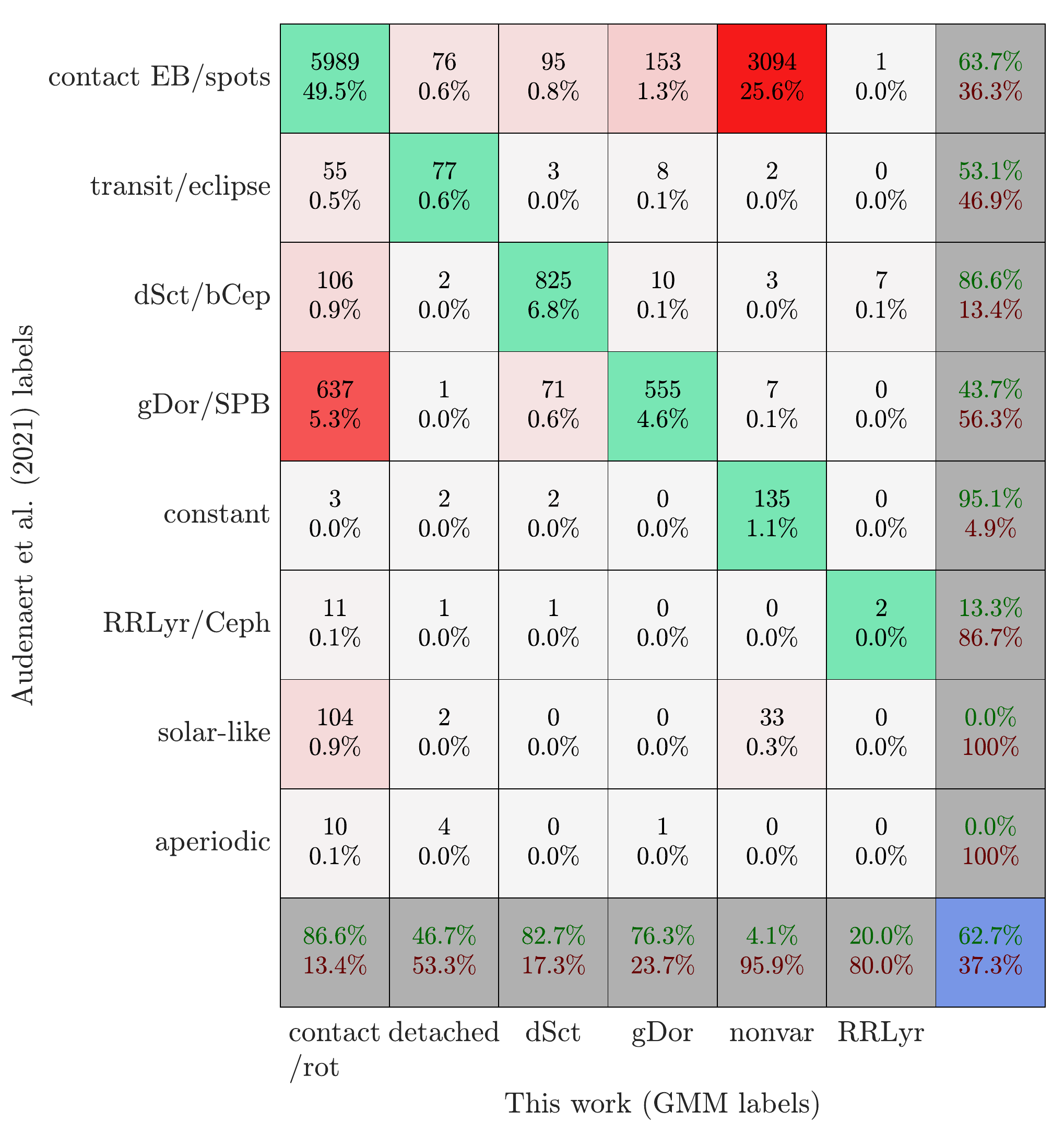}}
	\end{center}
	\caption{Confusion matrix comparing the results of the \citet{Audenaert++2021} classifier to our own classifications for about 12,000 stars in the {\kepler} field with $6500 \,\text{K} \le T_\mathrm{eff} \le 10,000 \,\text{K}$. Much like in Fig.~\ref{fig:confusion}, the grey summary boxes on the right correspond to the percentage of Aud21 labels that agree with our GMM predictions, while summaries at the bottom are the percentage of our predictions for each class that agree with Aud21.}
    \label{fig:confusion-audenaert}
\end{figure}

The classification by Aud21 included about 167,000 \kepler\ Q9 light curves, regardless of effective temperature, whereas our work is restricted to about 12,000 stars with $6500 \,\text{K} \le T_\mathrm{eff} \le 10,000 \,\text{K}$. We have compared our classifications with Aud21 in Fig.~\ref{fig:confusion-audenaert}. 
There is an obvious mapping between most of our classes and those used by Aud21, with the following differences:  
\begin{itemize}
    \item Aud21 combined contact eclipsing binaries and rotational (spotted) variables into a single class.
    \item Aud21 included \dsct\ stars in a class with $\beta$~Cephei stars.  These have similar light curves but the $\beta$~Cep pulsators have higher effective temperatures that lie outside the range of our sample.  Similarly, Aud21 combined \gdor\ stars with SPBs (Slowly Pulsating B stars), which are also hotter than our sample.
    \item Aud21 included a class for solar-like oscillators, which should not appear in our sample because they occur in stars whose effective temperatures fall below our range.
    \item Aud21 also included a class for aperiodic variables.
\end{itemize}

We see from the confusion matrix in Fig.~\ref{fig:confusion-audenaert} that there is generally excellent agreement between our results and those of \citet{Audenaert++2021}.  We briefly discuss the areas with the greatest disagreement:
\begin{enumerate}
        \item 3094 stars that our classifier labelled as non-variable were classified by Aud21 as rotational/contact EB.  We inspected 200 of these light curves (and their Fourier amplitude spectra) and found that most are non-variable, with some showing a weak rotation signal.  

        \item 637 stars that we labelled as contact EBs or rotational variables were classified by Aud21 as {\gdor} pulsators.  Inspection of 200 light curves shows that most are indeed \gdor\ stars.  This may be a shortcoming of our specific feature space and classifier, particularly when considering Fig.~\ref{fig:confusion}, where the same disagreement occurs between our GMM classifications and our independent test set.

        \item 153 stars were labelled by us as \gdor\ stars and by Aud21 as contact EBs or rotational variables.  Inspection of these shows that many are indeed \gdor\ stars, although it is sometimes difficult to be sure.

        \item 139 stars in our sample were labelled by Aud21 as having solar-like oscillations, which is not a class that we considered because these oscillations occur in stars below our temperature range.  Our classifications for these light curves were mainly as rotational variables, contact EBs or non-variable. We inspected all 139 light curves and found that our classifications were mostly correct. 

        \item 106 stars were labelled by us as contact EBs or rotational variables, and by Aud21 as \dsct\ stars.  We inspected all light curves and found that most have \dsct\ pulsations, but many also have low-frequency variability.
        
\end{enumerate}
Finally, we note KIC 10024862, which is one of two stars listed by Aud21 as non-variable and by our algorithm as a detached binary.  In fact, \citet{Kawahara+Kento2019} identified this as a Jupiter-sized exoplanet in a long-period orbit that has only one transit during the 4-year Kepler mission, which happened to be in Q9.  This suggests that it might be worthwhile to look in more detail at groups for which classification methods are in disagreement for a small number of stars.


\new{Much like our approach, Aud21 assigned labels to each star according to the class with the highest posterior probability from their classifier. Figure \ref{fig:confusion-audenaert} therefore contains samples where either classifier may be confused --- for example, a given light curve may have probabilities of 0.34, 0.35, 0.01 split between three classes and the maximum probability (0.35) is relatively low.  Not surprisingly, we found that by restricting to stars with a high maximum probability in both samples, the agreement increased between our classification labels and those from Aud21. A detailed comparison of the two catalogues goes beyond the scope of this paper and would require a measure of label confidence from the Aud21 classifier similar to the probability density heuristic from Section \ref{sec:pdf}.}

In general, we conclude that the two approaches produce results that generally agree well.  The difference in point (i) reflects the subjectivity in drawing the line between variables and non-variables (and perhaps also different amounts of filtering applied to the light curves).  Points (ii) and (iii) reflect the difficulty---especially with short data sets---in deciding whether low-frequency variability is due to pulsation or rotation \citep[e.g.,][]{Briquet++2007,Lee2021,Kurtz2022}.

\section{Conclusions}

We have used a feature-based machine learning algorithm to classify \kepler\ light curves for stars with effective temperatures in the range 6500--10,000\,K.  We first created a training set of 1319 light curves, which we classified into seven classes: $\delta$\,Scuti stars, $\gamma$\,Doradus stars, RR\,Lyrae stars, rotational variables, contact eclipsing binaries, detached eclipsing binaries, and non-variable stars. We built a classifier using features selected with the \hctsa\ package (highly comparative time-series analysis; \citealt{Fulcher+Jones2017}), which includes over 7000 time-series features.  We found that five features were sufficient to represent the training set with a balanced accuracy of 98\%, and a separate test set of 500 stars with a balanced accuracy of 82\%.

We used our method to classify \kepler\ light curves for all 12,000 stars with effective temperatures in the range 6500--10,000\,K, and the results are tabulated in the supplementary online material (Table~\ref{tab:posteriors}). We further outlined a confidence heuristic based on probability density with which to search our catalogue and extract candidate lists of correctly-classified variable stars. We also compared our classifications to recent work on the same light curves by \citet{Audenaert++2021} and generally found good agreement.

While many modern approaches to machine-learning focus on performance over interpretability (resulting in the common description of being `black box' algorithms), here we favoured the selection of high-performing and interpretable features with which to meaningfully represent {\kep} light curves.
Given the ease with which our five features can be computed for a large database of light curves, comparing complex classification algorithms to our methods could provide an independent benchmark for general light-curve classification algorithms, much like we have shown with our comparison to \cite{Audenaert++2021}.

Further extensions of this work might include using our catalogue to search for rare classes of variable stars, hybrid systems, and new stars entirely different to our training sample. In particular, we expect stars with roughly equal posterior probabilities between two classes to be hybrid systems, and very different stars to have much lower probability density scores than any other star in the {\kep} field. Our methods could also be applied to individual classes of variable stars to try to identify interesting or unusual behaviour within a class, such as the recently discovered high-frequency {\dsct} stars \citep{beddingetal2020}. Another possibility is to extend our intuitive feature-based methods by adding more complex feature selection and classification algorithms. Such extensions are likely to improve our already strong classification performance, and strengthen results when applying our methods to even larger photometric surveys, such as that from {\tess}.

\section*{Acknowledgements}

We thank the \kepler\ team for providing such a wonderful data set.
We gratefully acknowledge support from the Australian Research Council through DECRA grant  DE180101104 and Discovery Project DP210103119, and from the Danish National Research Foundation (Grant DNRF106) through its funding for the Stellar Astrophysics Centre (SAC).
TVR gratefully acknowledges support from the Research Foundation Flanders (FWO) under grant agreement number 12ZB620N.

\section*{Data Availability} 

The data underlying this article are available at the Kepler Asteroseismic Science Operations Center (KASOC), at \url{http://kasoc.phys.au.dk/}.  The \hctsa\ software is available at \url{https://github.com/benfulcher/hctsa}.


\input{output.bbl}


\bsp	
\label{lastpage}
\end{document}